\documentclass[prl,twocolumn,showpacs,superscriptaddress]{revtex4}
\usepackage{graphicx,bm,amssymb,amsmath}

\newcommand{\Eq}[1]{Eq.~\eqref{#1}}
\newcommand{\eq}[1]{\eqref{#1}}

\newcommand{\nn}{\nonumber}

\newcommand{\pdag}{{\phantom{\dagger}}}

\newcommand{\past}{{\phantom{\ast}}}

\DeclareMathOperator{\sgn}{sign}
\DeclareMathOperator{\tr}{Tr}

\newcommand{\PRL}[3]{Phys. Rev. Lett.~\textbf{#1}, #2 (#3)}

\newcommand{\PRA}[3]{Phys. Rev. A~\textbf{#1}, #2 (#3)}
\newcommand{\Science}[3]{Science~\textbf{#1}, #2 (#3)}
\newcommand{\Nature}[3]{Nature~\textbf{#1}, #2 (#3)}
\newcommand{\JMP}[3]{J. Math. Phys.~\textbf{#1}, #2 (#3)}
\newcommand{\JSM}[2]{J. Stat. Mech.~#1 (#2)}

\newcommand{\etal}{\textit{et al.}}
\begin{document}

\title{Correlations in an expanding gas of hard-core bosons}

\author{D.M. Gangardt}
\affiliation{
School of Physics and Astronomy, University of Birmingham,
Edgbaston, Birmingham B15 2TT, UK
}
\author{M. Pustilnik}
\affiliation{
School of Physics, Georgia Institute of Technology,
Atlanta, GA 30332, USA
}

%\date{\today}

\begin{abstract}
We consider a longitudinal expansion of a one-dimensional gas of hard-core
bosons suddenly released from a trap. We show that the broken translational 
invariance in the initial state of the system is encoded in correlations between 
the bosonic occupation numbers in the momentum space. The correlations 
are protected by the integrability and exhibit no relaxation during the expansion. 
\end{abstract}

\maketitle

Rapid progress in the ability to manipulate ultracold atomic gases stimulated
a revival of interest in fundamental properties of interacting  Bose systems. 
Unlike conventional condensed matter systems, cold gases offer a unique 
possibility to monitor the out-of-equilibrium dynamics of interacting 
systems unhindered by coupling to the environment and associated with it 
decoherence, see~\cite{Bloch} for a recent review. Among various realizations 
of interacting Bose systems, the one-dimensional (1D) 
ones~\cite{hardcore-exp,Weiss_cradle,Bragg_1D} 
occupy a special place: interactions in 1D have a much stronger effect than in 
higher dimensions while often allowing for a complete theoretical treatment.

This paper is partially motivated by the recent experimental study~\cite{Weiss_cradle}
of the evolution of 1D strongly interacting Bose liquid from a carefully prepared
nonequilibrium initial state. The experiment~\cite{Weiss_cradle} showed that
the momentum distribution function does not exhibit a noticeable relaxation 
towards equilibrium. Such extremely slow relaxation is consistent with the 
behavior expected for an almost integrable system. Indeed, in a 1D system 
only three-particle collisions may lead to a momentum relaxation; such processes 
are absent in integrable models~\cite{Sutherland}.

On the theoretical side, experiments such as~\cite{hardcore-exp,Weiss_cradle} 
highlight the relevance of the well-known in statistical mechanics 
\textit{quantum quench} problem: how to describe the evolution 
of a system from an arbitrary initial state (see, e.g.,~\cite{Cardy} 
and references therein). At present, there are very few exact results 
on such strongly nonequilibrium dynamics of interacting quantum systems. 
Various problems of this type arise naturally in the description of the 
experiments on trapped cold atomic gases. Indeed, by far the most 
popular technique today is to observe the expansion of a gas after a sudden 
release of the trap~\cite{Bloch}. Although such experiments are obviously 
destructive, the time-of-flight imaging~\cite{Bloch} allows one to study the 
real-time evolution of the bosonic occupation numbers in the momentum space. 
Importantly, not only the average occupation numbers (momentum distribution) 
but also the corresponding higher-order statistical moments (fluctuations) are 
accessible experimentally~\cite{Altman,noise_exp}). Unlike the momentum 
distribution, the fluctuations are sensitive to the relaxation in the system. 

We consider a simple yet realistic~\cite{hardcore-exp} example: 
expansion of a 1D gas of bosons with infinitely strong contact repulsion 
(hard-core bosons) suddenly released from a trap. We show that shortly after 
the trap release, bosonic occupation numbers reach their steady-state values. 
We derive an operator identity, see \Eq{20} below, that relates the bosonic 
occupation numbers in the steady state to the integrals of motion. The identity 
allows one to study all statistical moments of the bosonic occupation numbers, 
and we evaluate the second moment in a closed form. Correlations between the 
occupation numbers at different momenta reflect directly the broken translational 
invariance in the initial (trapped) state of the system.

To be specific, we assume that initially (at $t<0$) the system is in a thermal
equilibrium state of the Hamiltonian
\begin{equation}
H = H_0 + V_\text{trap}.
\label{2}
\end{equation}
Here $H_0$ describes 1D hard-core Bose gas without confinement 
(see \Eq{5} below) and
\begin{equation}
V_\text{trap} = \int\!dx\,V(x)\rho(x)
\label{3}
\end{equation}
describes the effect of the trap, with $\rho(x)$ being the local density
operator. The trap potential \eq{3} breaks translational invariance, hence
it does not commute with the Hamiltonian $H_0$ that governs the dynamics
after the trap release at $t=0$.

The simplest description of the hard-core bosons is based on the
Jordan-Wigner transformation
\begin{equation}
\psi(x) = \exp\!\left[i\pi\!\int^x_{-\infty}\!\!dy\, \rho(y)\right]\varphi(x),
\label{4}
\end{equation}
where the operators $\psi$ and $\varphi$ correspond to fermions
and bosons, respectively:
\[
\bigl\{\psi(x),\psi^\dagger(y)\bigr\}
= \bigl[\varphi(x),\varphi^\dagger(y)\bigr]
= \delta (x-y).
\]
The transformation \eq{4} maps the hard-core bosons onto free spinless
fermions~\cite{Girardeau},
\begin{equation}
H_0 = \int\!dx\,\psi^\dagger (x) 
\left[- \frac{1}{2m}\frac{\partial^2}{\partial x^2}\right]
\psi(x).
\label{5}
\end{equation}
At the same time, the local density operator [and, therefore, \Eq{3}],
retains its form,
\[
\rho(x) = \psi^\dagger(x)\psi(x) = \varphi^\dagger(x)\varphi(x).
\]

Since the fermionic occupation numbers in the momentum space
\begin{equation}
n_k = \psi^\dagger_k\psi^\pdag_k,
\quad
\psi^\pdag_k = (2\pi)^{-1/2}\!\!\int\!dx\,e^{-ikx}\psi(x),
\label{6}
\end{equation}
commute with $H_0$, the expectation values of $n_k$ are independent of time,
\begin{equation}
\langle n_k\rangle^\pdag_t = \langle n_k\rangle^\pdag_0 = \text{const}.
\label{7}
\end{equation}
Hereinafter
\[
\langle \widehat{\mathcal O}\,\rangle^\pdag_t
= \langle e^{iH_0t}\widehat{\mathcal O}\,e^{-iH_0t}\rangle^\pdag_0,
\]
where
$\langle \ldots\rangle_0 \equiv \langle \ldots\rangle^\pdag_{t\to 0}$
denotes the thermal averaging with the initial Hamiltonian
$H$, see \Eq{2}.

The crucial for the following observation is that even though the evolution
of the system at $t>0$ is governed by the \textit{translationally invariant}
Hamiltonian $H_0$, the initial Hamiltonian $H$ does not have this symmetry.
Therefore, not only the diagonal in $k$ fermionic bilinears (such as $n_k$)
have finite expectation values, but also the off-diagonal ones, e.g.,
$\langle\psi^\dagger_k\psi^\pdag_{k'}\rangle^\pdag_0\neq 0$. 
Since the expectation value
\begin{equation}
\langle\psi^\dagger_k\psi^\pdag_{k'\!}\rangle^\pdag_t
= \langle\psi^\dagger_k\psi^\pdag_{k'\!}\rangle^\pdag_0\,
e^{i(\epsilon_k - \epsilon_{k'\!})\,t},
\quad
\epsilon_p = p^2\!/2m,
\label{8}
\end{equation}
oscillates with $t$, quantities such as
$\bigl|\langle\psi^\dagger_k\psi^\pdag_{k'}\rangle^\pdag_t\bigr|$
remain constant and carry with them the memory of the broken-symmetry 
initial state of the system. Because the off-diagonal correlation functions 
\eq{8} are finite, the fermionic occupation numbers fluctuate. Indeed, 
with the help of the Wick theorem one finds for 
$\delta n_k = n_k - \langle n_k\rangle^\pdag_0$
\begin{equation}
\langle \delta n_k \delta n_{k'\!}\rangle^\pdag_t
= - \bigl|\langle\psi^\dagger_k\psi^\pdag_{k'\!}\rangle^\pdag_0\bigr|^2
= \text{const}.
\label{9}
\end{equation}

It should be emphasized that the very survival of quantities such as Eqs. \eq{8}
and \eq{9} that preserve the information about the initial conditions does
not rely on the particularly simple form that the Hamiltonian $H_0$ has
in our case. Rather, it is a direct consequence of the integrability. Indeed,
in a generic (nonintegrable) system correlation function \eq{8} would
decay with $t$. This decay (relaxation) ``washes out'' the memory about 
the symmetry of the initial state, thereby restoring the translational invariance. 
After the relaxation is complete, the density matrix commutes with the total 
momentum. Averaging with any density matrix that has this symmetry would 
give zero for the correlation functions such as \Eq{8}.

A very similar consideration can be applied to any quantum quench problem
in which the symmetry of the Hamiltonian that governs the system's dynamics
differs from that of the initial state. [For example, suddenly turned off 
interactions in the Luttinger model~\cite{Cazalilla} correspond to the initial 
state with broken global U$(1)$ symmetry]. The information about the symmetry
of the initial state is encoded in the off-diagonal correlation functions [cf. \Eq{8}];
relaxation manifests itself in the decay of these off-diagonal correlations with time. 

We now consider a specific but rather realistic situation when the trap potential
\Eq{3} is harmonic,
\begin{equation}
V(x) =\frac{1}{2}\, m\omega^2 x^2
= \frac{x^2}{2 m l^4}\,,
\quad
l = (m\omega)^{-1/2}.
\label{10}
\end{equation}
At zero temperature the correlation function \eq{8} can be written as
\begin{equation}
\langle\psi^\dagger_k\psi^\pdag_{k'\!}\rangle^\pdag_0
= \sum_{n=0}^{N-1} \phi_n^\ast(k)\phi_n^\past\!(k'),
\label{11}
\end{equation}
where $N\gg 1$ is the number of particles in the system and $\phi_n(k)$ is
the stationary eigenfunction in the momentum representation that corresponds
to $n$-th energy level of a harmonic oscillator. The expectation values of the 
fermionic occupation numbers \eq{7} are obtained by setting $k=k'$ in \Eq{11}. 
For $N\gg 1$, this yields a ``semicircle'' dependence
\begin{equation}
\langle n_k \rangle^\pdag_0
=  \frac{R}{\pi}\sqrt{1-k^2\!/k_F^2}\,,
\label{12}
\end{equation}
where $k_F$ is the Fermi momentum and $R$ is the classical radius of
$N$-particle fermionic cloud confined in a harmonic trap, 
\[
k_F l = R/l = \sqrt{2N}.
\]

In writing \Eq{12} we neglected the oscillating with $k$ contribution that
has a relative magnitude of the order of $l/R\ll 1$. This
contribution is the momentum-space counterpart of the Friedel oscillations
in $\langle \rho(x)\rangle^\pdag_0$, as it is obvious from the operator
identity~\cite{identity}
\begin{equation}
n_k = \,e^{iH\tau}\bigl[ l^2\!\rho(k l^2)\bigr] \,e^{-iH\tau} ,
\quad
\tau = \frac{\pi}{2\omega}\,.
\label{13}
\end{equation}
The period of the Friedel oscillations in $x$-space is the Fermi wavelength
$2\pi/k_F$. \Eq{13} then implies that the corresponding oscillations in
$k$-space have a period $2\pi/k_F l^2 = 2\pi/R$.

Although the Friedel oscillations contribute little to $\langle n_k \rangle^\pdag_0$,
they are responsible for the fluctuations of the fermionic occupation numbers.
Indeed, using Eqs. \eq{9} and \eq{11}, we find 
\begin{equation}
\langle \delta n_k \delta n_{k'\!}\rangle^\pdag_0
= - \,\frac{\sin^2\bigl[(k-k')R\bigr]}{\pi^2(k-k')^2}\,.
\label{14}
\end{equation}
\Eq{14} is valid when both $|k|$ and $|k'|$ are small compared with the
Fermi momentum $k_F$. In this limit the dependence of
$\langle \delta n_k \delta n_{k'\!}\rangle^\pdag_0$ on $k+k'$
[which we have neglected in writing \Eq{14}] is very weak.

The visibility of the Friedel oscillations \Eq{14} is not affected by 
temperature $T$ as long as $T\ll \epsilon_F$, where 
$\epsilon_F= N\omega = k_F^2\!/2m$ is the Fermi energy 
(i.e., the chemical potential for hard-core bosons).
Indeed, a finite temperature introduces an uncertainty $\delta R\sim RT/\epsilon_F$ 
in the size of the cloud which leads to the exponential decay of
$\langle \delta n_k \delta n_{k'\!}\rangle^\pdag_0$ at $|k-k'|\gtrsim 1/\delta R$.
The oscillations \eq{14} survive as long as $\delta R\ll R$, i.e., at 
$T\ll \epsilon_F$.

So far, we demonstrated that the information about the broken translational
invariance in the initial state of the system is preserved in the statistics of the
\textit{fermionic} occupation numbers. In particular, it manifests itself in the
characteristic oscillatory dependence of
$\langle \delta n_k \delta n_{k'\!}\rangle^\pdag_0$ on $k-k'$,
see \Eq{14}. However, the fermions emerged in our problem merely as a 
convenient way of dealing with the exact eigenstates of the system. Since 
the relation between the effective fermions and the original hard-core bosons 
is nonlocal, see \Eq{4}, the behavior of the \textit{bosonic} correlation 
functions is much more complex.

We discuss here the bosonic occupation numbers
$f_k^\pdag\! = \varphi_k^\dagger\varphi_k^\pdag$. 
Unlike their fermionic counterparts Eqs. \eq{7} and \eq{9}, 
the expectation value $\langle f_k\rangle^\pdag_t$
(momentum distribution) and the fluctuations 
$\langle \delta f_k \delta f_{k'\!}\rangle^\pdag_t$
(here $\delta f_k = f_k - \langle f_k\rangle^\pdag_t$)
are no longer constant. However, the time-dependent contributions 
are superpositions of an infinite number of oscillating terms and 
decay at $t\to\infty$~\cite{decay}.

In order to find the bosonic occupations at $t\to\infty$, we again
concentrate on the harmonic trap potential \Eq{10}.
Following the method of~\cite{MG,BS,Kagan}, we consider first
the single-particle Schr\"odinger equation
\[
i\,\frac{\partial}{\partial t}\,\psi_n(x,t) 
= - \frac{1}{2m}\frac{\partial^2}{\partial x^2}\,\psi_n(x,t)
\] 
with the initial condition
$\psi_n(x,0)=\phi_n(x)$, where $\phi_n(x)$ is the normalized stationary eigenfunction
of a harmonic oscillator corresponding to the eigenenergy 
$\epsilon_n = (n+1/2)\omega$. At $t\gg 1/\omega$,
the wave function $\psi_n(x,t)$ assumes the form~\cite{Perelomov}
\begin{equation}
\psi_n(x,t) =
\frac{1}{\sqrt{\omega t}}\,\exp\left(\frac{\,i x^2}{2l^2\omega t}\right)
e^{-i\epsilon_n\tau}\phi_n\!\left(x/\omega t\right),
\label{50}
\end{equation}
where $\tau$ is given by \Eq{13}. Upon introducing dimensionless variables
\[
\eta = \omega t,
\quad 
\xi = \frac{x}{\eta l}\,,
\]
we rewrite \Eq{50} as
\begin{equation}
\psi_n(x,t) = \eta^{-1/2}e^{i\eta \xi^2\!/2}\, e^{-i\epsilon_n\tau}\phi_n\!\left(\xi l\right).
\label{15}
\end{equation}
Using \Eq{15}, the first-quantized many-particle wave function of
hard-core bosons~\cite{hardcore} $\Phi_t$ can be expressed via
its initial value $\Phi_0$,
\begin{equation}
\Phi_t\bigl(\{x_i\}\bigr) = \eta^{-N/2}
e^{i\eta\!\sum \xi_i^2\!/2} e^{-iE_0\tau}\Phi_0\bigl(\{\xi_i l\}\bigr) ;
\label{16}
\end{equation}
here $\{x_i\} = x_1,\ldots,x_N$, $\xi_i = x_i/\eta l$, and 
$\Phi_0\bigl(\{x_i\}\bigr)$ is the many-body eigenstate of the initial 
Hamiltonian $H$ with energy $E_0$.
 
In the second-quantized language, \Eq{16} implies the operator
relation~\cite{identity}
\begin{equation}
\varphi(x,t)
 = \eta^{-1/2}e^{i\eta \xi^2\!/2} \widetilde\varphi(\xi l,\tau),
\label{17}
\end{equation}
where $\varphi(x,t)$ and $\widetilde\varphi(x,t)$ are operators in the
Heisenberg representation with the time dependence governed by
the Hamiltonians $H_0$ and $H= H_0+V$, respectively:
\[
\varphi(x,t) = e^{i H_0 t} \varphi (x) \,e^{- i H_0 t},
\quad
\widetilde\varphi(x,t) = e^{i H t} \varphi (x) \,e^{- i H t}.
\]
Substitution of \Eq{17} into
\[
f_k(t) 
= \frac{1}{2\pi}\!
\int\!dx\,dx'\, e^{ik(x-x')} \varphi^\dagger(x,t)\varphi(x',t)
\]
yields
\begin{eqnarray}
f_k(t) &=& \frac{l^2\eta}{2\pi}\!\int\!d\xi\,d\xi'e^{i\eta(\xi-\xi')\bigl[k l-(\xi+\xi')/2\bigr]\,}
\nn\\
&&\qquad\quad\times\,\widetilde\varphi^\dagger(\xi l,\tau)\widetilde\varphi(\xi'l,\tau).
\label{18}
\end{eqnarray}
At $\eta\to\infty$ the integral over $\xi$ and $\xi'$ here can be evaluated
in the stationary phase approximation with the result~\cite{identity}
\begin{equation}
f_k(t\to\infty) = e^{iH\tau}\bigl[l^2\!\rho\bigl(k l^2\bigr)\bigr]e^{-iH\tau}.
\label{19}
\end{equation}
[Analogous calculation for fermions yields \Eq{13} which,
unlike \Eq{19}, is valid at all $t>0$.]
Finally, comparing \Eq{19} with \Eq{13}, we find~\cite{identity}
\begin{equation}
f_k(t\to\infty) = n_k.
\label{20}
\end{equation}

According to \Eq{20}, at $t\to\infty$ the bosonic occupation numbers 
in $k$-space $f_k$ coincide with the integrals of motion $n_k$. Since 
\Eq{20} holds for operators~\cite{identity}, it also implies that the 
statistical moments of the bosonic occupation numbers $f_k$ at $t\to\infty$ 
coincide with those for fermions in the initial trapped state, e.g.,
\begin{equation}
\langle f_k\rangle^\pdag_\infty=\langle n_k\rangle^\pdag_0,
\quad
\langle \delta f_k \delta f_{k'\!}\rangle^\pdag_\infty
= \langle \delta n_k \delta n_{k'\!}\rangle^\pdag_0.
\label{100}
\end{equation}

We pause now to discuss the conditions of applicability of Eqs. \eq{20} 
and \eq{100}. Since we used the stationary phase method, the relevant 
characteristic time scales can be obtained by equating the scale of 
variation with $\xi$ of the phase $\eta\xi^2$ in \Eq{17} with that 
of the field $\widetilde\varphi(\xi l,\tau)$. 

The dependence of $\varphi(x)$ [and, therefore, of $\widetilde\varphi(x,\tau)$] 
on $x$ is characterized by two length scales. The longer one is the
size of the trapped system $R$. Neglecting all other scales, we find 
$t_1\sim 1/\epsilon_F$; this corresponds to the time the particles in the 
trap move between the collisions (the mean free time). The shorter scale of 
variation of $\varphi(x)$ is the distance between particles $R/N$, which 
leads to the time scale $t_2\sim 1/\omega\sim Nt_1$. This is the time it takes 
for a particle moving with the Fermi velocity $v_F = k_F/m$ to cross the trap, 
$t_2\sim R/v_F$, or, equivalently, for the expanding cloud to double its size.

Shortly after release of the trap, at $t_1\ll t\ll t_2$, the oscillating 
transient contributions to the bosonic momentum distribution are 
still present, but $\langle f_k\rangle^\pdag_t$ averaged over time 
is already given by the smooth ``fermionic'' semicircle \Eq{12}, 
see \cite{MG}. In this regime the discreteness of the system is not 
important and the ``shot noise'' fluctuations, \Eq{14}, are not yet resolved. 
Accordingly, the \textit{hydrodynamic} description~\cite{MG,BS} based 
on \Eq{12} provides a complete information about the system.

Much later, at $t\gg t_2$, the system enters the asymptotic regime 
where \Eq{20} is applicable. In this regime the transients have already 
decayed~\cite{decay}, the statistics of particles no longer matters,
and statistical moments of the bosonic occupation numbers approach 
their steady-state values \Eq{100}. In other words, this regime is 
essentially that of the \textit{collisionless expansion} of the system.

It should be noted that the setup discussed here is essentially the same 
as that studied recently in~\cite{Olshanii}. Based on the behavior of 
$\langle f_k\rangle^\pdag_t$, it was conjectured there that 
any isolated system with integrable dynamics \textit{relaxes} 
to a state described by a certain generalized Gibbs distribution. 
According to the prescription 
of~\cite{Olshanii} adopted for continuously varying $k$~\cite{finite}, 
the density matrix at $t\to\infty$ has the form
$\hat\rho_G\propto\exp\left(-\int\!dk\,\beta_k n_k\right)$.
Although the coefficients $\beta_k$ can always be chosen in such a way that 
$\tr(\hat\rho_G n_k)= \langle n_k\rangle^\pdag_0$~\cite{Olshanii}, 
finding $\beta_k$ is obviously not sufficient to establish the validity of the 
conjecture. Indeed, no matter what $\beta_k$ is,
$\langle n_k n_{k'}\rangle
= \tr(\hat\rho_G n_k n_{k'}\!)
\equiv \langle n_k\rangle\langle n_{k'}\rangle$, 
i.e., $\langle \delta n_k \delta n_{k'\!}\rangle = 0$.
In view of Eqs. \eq{14}, \eq{20}, and \eq{100}, we conclude that
$\hat\rho_G$ not only neglects the correlations between different integrals 
of motion $n_k$, but also fails to account for the correlations between the 
bosonic occupation numbers $f_k$. 
This raises serious doubts whether the generalized Gibbs distribution 
conjectured in~\cite{Olshanii} is actually useful
for the description of the quantum quench problems.

Although our consideration relied rather heavily on the properties of the hard-core
Bose gas with the harmonic initial confinement, we expect some of our conclusions
to be generic. In particular, we expect that any finite 1D system with short-range
interactions enters the collisionless expansion regime 
at $t\gg t_2\sim R/v_s$, where $v_s$ is the sound velocity in the 
initial trapped state. 
(In an infinite system, this time scale corresponds to the establishment 
of a local equilibrium in a subsystem of size $R$~\cite{Cardy}). 

In a nonintegrable system, a relaxation would occur at $t\lesssim t_2$. 
The relaxation would partially restore the translational invariance,
leading to the suppression of the noise 
$\langle \delta f_k\delta f_{k'}\rangle^\pdag_\infty$. Since the noise 
is not sensitive to temperature (see above), the accuracy of noise 
measurements in time-of-flight experiments is not limited by one's 
ability to control the temperature. This suggests that deviations from 
the integrability are easier to detect in noise measurements than, for 
example, by observing the saturation of the height of the peak in the 
dynamic structure factor with lowering the temperature~\cite{Khodas} 
in Bragg spectroscopy experiments~\cite{Bragg_1D}. 

Finally, we emphasize that real-life 1D bosons are neither hard core
nor their dynamics is integrable. It is conceivable that deviations
from integrability will have a dramatic effect on the behavior of some
observable quantities. Detailed understanding of the relaxation mechanisms 
and other consequences of nonintegrability in 1D Bose systems
remains an important open problem.

\begin{acknowledgments}
We thank Natan Andrei and Maxim Olshanii for valuable discussions
and Kavli Institute for Theoretical Physics at UCSB and Abdus
Salam ICTP for the hospitality. 
This project is supported by
EPSRC Advanced Fellowship EP/D072514/1 (DMG), 
and by NSF grants DMR-0604107 (MP)
and PHY05-51164.
\end{acknowledgments}


\vspace{-0.05in}
\begin{thebibliography}{99}
\vspace{-0.05in}

\bibitem{Bloch}
I. Bloch, J. Dalibard, and W. Zwerger, arXiv:0704.3011.
\bibitem{hardcore-exp}
B.  Paredes \etal, 
%Widera A  , Murg V  , Mandel O  , Folling S  , Cirac I  , Shlyapnikov GV  , Hansch TW  , Bloch I  
\Nature{429}{277}{2004};
T. Kinoshita, T. Wenger, and D.S. Weiss, 
\Science{305}{1125}{2004}.
\bibitem{Weiss_cradle}
T. Kinoshita, T. Wenger, and D.S. Weiss, 
\Nature{440}{900}{2006}.
\bibitem{Bragg_1D}
T. St\"oferle \etal, \PRL{92}{130403}{2004}.
\bibitem{Sutherland}
B. Sutherland, \textit{Beautiful Models} (World Scientific,
Singapore, 2004).
\bibitem{Cardy}
P. Calabrese and J. Cardy,  \JSM{P06008}{2007}.
\bibitem{Altman}
 E. Altman, E. Demler, and M.D. Lukin, \PRA{70}{013603}{2004}.
\bibitem{noise_exp}
M. Greiner \etal, %C.A. Regal, J.T. Stewart, and D.S. Jin,
\PRL{94}{110401}{2005};
S.~F\"olling \etal, %F. Gerbier, A. Widera, O. Mandel, T. Gericke, and I. Bloch,
\Nature{434}{481}{2005}.
\bibitem{Girardeau}
M.D. Girardeau, \JMP{1}{516}{1960}.
\bibitem{Cazalilla}
M.A. Cazalilla, \PRL{97}{156403}{2006}.
\bibitem{identity}
The operator relations Eqs. \eq{13}, \eq{17}, \eq{19}, and \eq{20} hold
in the sense that the left- and the right-hand sides have the same matrix elements
between any two many-particle eigenstates of the Hamiltonian \eq{2} with 
harmonic confinement \eq{3}, \eq{10}.
\bibitem{decay}
The decay of transients is analogous to the spreading of a wave packet in
the single-particle quantum mechanics and should not be confused
with relaxation.
\bibitem{MG}
A. Minguzzi and D.M. Gangardt, \PRL{94}{240404}{2005}.
\bibitem{BS}
B. Sutherland, \PRL{80}{3678}{1998}.
\bibitem{Kagan}
Yu. Kagan, E.L. Surkov, and G.V. Shlyapnikov,
\PRA{54}{R1753}{1996}.
\bibitem{Perelomov}
A.M. Perelomov and Ya.B. Zel'dovich,
\textit{Quantum Mechanics: Selected Topics}
(World Scientific, Singapore, 1999).
\bibitem{hardcore}
The many-particle wave function of hard-core bosons is given by
$
\Phi_t\bigl(\{x_i\}\bigr) = \Psi_t\bigl(\{x_i\}\bigr) \prod_{i<j}\sgn(x_i-x_j).
$
Here $\Psi_t$ is the %\textit{fermionic} wave function given by a 
Slater determinant built of the single-particle wave functions
$\psi_n(x,t)$~\cite{Girardeau}.
\bibitem{Olshanii}
M. Rigol \etal, % V. Dunjko, V. Yurovsky, and M. Olshanii,
\PRL{98}{050405}{2007};
M.~Rigol, A. Muramatsu, and M. Olshanii,
\PRA{74}{053616}{2006}.
\bibitem{finite}
The results \eq{13}, \eq{14}, \eq{100} do not rely on the continuity of $k$ 
and remain intact when the system expands in a finite box 
of the size $L\gg R$ rather than in a vacuum.
\bibitem{Khodas}
M. Khodas \etal, \PRL{99}{110405}{2007}.

\end{thebibliography}
\end{document}